\documentclass[fleqn,10pt]{wlscirep}
\usepackage{amsmath,chemarrow}
\usepackage{color}
\title{Enhanced robustness of evolving open systems by the bidirectionality of interactions between elements}

\author[1,4,6*]{Fumiko Ogushi}
\author[2,3,4]{J\'{a}nos Kert\'{e}sz}
\author[4]{Kimmo Kaski}
\author[5,4$\dagger$]{Takashi Shimada}
\affil[1]{Advanced Institute for Material Research, Tohoku University, 2-1-1 Katahira, Aoba-ku, Sendai, 980-8577, JAPAN}
\affil[2]{Center for Network Science, Central European University, 1051 Budapest, Hungary}
\affil[3]{Institute of Physics, Budapest University of Technology and Economics, 1111 Budapest, Hungary}
\affil[4]{Department of Computer Science, Aalto University School of Science, P.O. Box 15500, Espoo, Finland}
\affil[5]{Quantum-Phase Electronics Center and Department of Applied Physics, Graduate School of Engineering, The University of Tokyo, 7-3-1 Hongo, Bunkyo-ku, Tokyo, 113-8656, JAPAN}
\affil[6]{Center for Materials research by Information Integration, National Institute for Materials Science, 1-2-1 Sengen, Tsukuba, Ibaraki 305-0047, JAPAN}
\affil[*]{fumiko.ogushi.e3@tohoku.ac.jp}
\affil[$\dagger$]{shimada@ap.t.u-tokyo.ac.jp}

\begin{abstract}
Living organisms, ecosystems, and social systems are examples of complex systems in which robustness against inclusion of new elements is an essential feature. 
A recently proposed simple model has revealed a general mechanism by which such systems can become 
robust against inclusion of elements with random interactions when the elements have a moderate number of links.
This happens as a result of two opposing effects such that while the inclusion of elements with more interactions makes each individual element more robust against disturbances, it also increases the net impact of the loss of any element in the system.
The interaction is, however, in many systems often intrinsically bidirectional like for mutual symbiosis, competition in ecology, and the action-reaction law of Newtonian mechanics, etc. 
This study reports the strong reinforcement effect of the bidirectionality of the interactions on the robustness of evolving systems. We show that the system with purely bidirectional interactions can grow with two-fold average degree, in comparison with the purely unidirectional system. 
This drastic shift of the transition point comes from the reinforcement of each node, not from a change in structure of the emergent system. For systems with partially bidirectional interactions we find that the area of the growing phase gets expanded.
In the dense interaction regime, there exists an optimum proportion of bidirectional interactions for the growth rate at around $1/3$.  
In the sparsely connected systems, small but finite fraction of bidirectional links can change the system's behaviour from non-growing to growing behaviour. 
\end{abstract}

\begin{document}

\flushbottom
\maketitle
%
%
\thispagestyle{empty}


\section*{Introduction}
An important and universal feature of many social, economical, ecological, and biological systems is that they are open. In these complex systems the constituting elements are not fixed and the complexity emerges or at least persists under successive appearances or introductions of new elements and disappearances or eliminations of old elements. Those systems sometimes grow or are stationary, but also some other times they collapse or go extinct. Hence one can ask a fundamental question why and when, in general, can such open and complex systems exist 
\cite{Gardner1970, Pimm1979, BakSneppen1993, Tokita1999, AlbertJeongBarabasi2000, Christensen2002, Jain2002, Kondoh2003, MoreiraJoseHans2009, Juan2009, BuldyrevStanleyHavlin2010, JanNagler2014, Yakubo2015}.

In the previous studies, we have revisited this classical problem using a very simple graph dynamics model.
In this model, the system is composed of a collection of nodes connected by uni-directional links with weights.
The nodes may represent various kinds of constitutive elements of the system i.e. chemicals, genes, animals, individuals, or species of some sort as we will term them from now on. Here the links are assumed to describe different types of directed influences between pairs of species. 
The strength of the influence of species $j$ on species $i$ is denoted by the weight of the uni-directional link from node $j$ to node $i$, $a_{ij}$. The weights can be either positive or negative. Each species has only one property that determines the dynamics, namely ``fitness'', which is simply given by the sum of its incoming interactions from other species in the system, i.e., $f_i = \sum^{in}_j a_{ij}$. A species can survive as long as its fitness is greater than zero; otherwise, it goes extinct. We calculate the fitness for each species and identify the species with minimum fitness and if it turns out to be non-positive, it will be deleted. Because each extinction event will also modify the fitness of the other species, the fitness is re-calculated and the least-fit species is re-identified. The deletion of species is continued until the minimum fitness value of a species becomes positive, meaning that nothing more will happen in terms of the above process. 
After finding such a state, we proceed to the next time step by adding a new species into the system. The $m$ interacting species are chosen randomly from the resident species with equal probability and the directions are also determined randomly. The link weights are again assigned randomly using the standard normal distribution.
Then, we re-calculate the fitness of each species to find the species that will become extinct.

The results of the model have shown that the repetition of this process gives rise either to continuous growth or stagnation for the system size, depending on the model's unique parameter, namely the number of weighted uni-directional links per a new node, $m$. 
Furthermore it turns out that the system can grow only if the connections are moderately sparse, i.e. the average degree is within a rather narrow interval for $5 \le m \le 18$. If the network becomes denser, there is a transition from growth to decay. It was shown that this transition originates from a balance between two effects, namely while on one hand the inclusion of species with more interactions makes each node more robust, on the other hand it also increases the impact of the loss of a node\cite{Shimada2014SREP, Shimada2015MABS}. This relation might be the origin of the moderately sparse network structure (with average degree up to few tens\cite{Winemiller1989,BA_RevModPhys2002, Ma_GRNsparseness2007, JANE:JANE1460}) ubiquitously found in real world cases, and the non-trivial distribution function of the lifetime of species in the system\cite{Murase2010}.

In the previous model, the interactions between species of the system were generated randomly with certain probability as unidirectional links between a pair of species $i$ and $j$. This means that a random unidirectional link from $j$ to $i$ with a unidirectional link from $i$ to $j$ existing already, could be generated as a second order effect proportionally to its link density, as $\sim \rho^2$. Hence the existence of influence from species $i$ to $j$ is independent of the existence of an influence in the opposite direction, i.e. from $j$ to $i$. However, the simulation results of this model of self-emergence have shown that the accumulation of bidirectional influence is negligible, especially for large systems their density turned out to be vanishingly small. 

However, in real systems, the interactions can on one hand be considered unidirectional as in case of reaction dynamics or signaling, and on the other hand bidirectional reflecting various types of bidirectional influence, like symbiosis or competition between species, predation-prey ecology, and even action-reaction law of Newtonian mechanics, etc.
Therefore, in real systems we could distinguish five main types of interactions, of which two are unidirectional in nature, and three reflect bidirectional influence in terms of being cooperative, predatory, and competitive, as shown in Table\ref{table_InteractionClasses} with the characterization and some real life examples. 

Our aim in this study is to gain better understanding of the robustness and growth of real systems, with computational modeling. In this study we will focus on studying systems in which the interactions around the species newly introduced into a system can be unidirectional, bidirectional, or a mixture of both in certain proportions. In the Model section we revisit the previous model and extend it to include bidirectional influence between species. In the following Results section we describe the simulation results for the cases purely and partially bidirectional interactions. Next in the Discussion section we summarize our findings. This is then followed with the extensive Methods section. 

\begin{table}[b]
\centering 
\begin{tabular}{|l|c|l|}
	\hline
	interaction type & representation & examples in real systems\\
	\hline
    Excitatory & $\displaystyle X \autorightarrow{$+$}{} Y$ &
    signaling in neuronal and gene regulatory nets, commensalism, donation\\
    \hline
    Inhibitory & $\displaystyle X \autorightarrow{$-$}{} Y$ & signaling in neuronal and gene regulatory nets, amensalism, harassment\\
	\hline
	Cooperative & $\displaystyle X \autorightleftharpoons{$+$}{$+$} Y$ & positive feedback, mutualism, cooperation, friendship\\
	\hline
	Predatory & $\displaystyle X \autorightleftharpoons{$-$}{$+$} Y$ &  negative feedback, predation, parasitism, exploitation, selfish actions\\
	\hline
	Competitive & $\displaystyle X \autorightleftharpoons{$-$}{$-$} Y$ & mutual inhibition, competition, antagonism\\
	\hline
\end{tabular}
\caption{Types of interactions with those representation by links, and examples in real systems.}
\label{table_InteractionClasses}
\end{table}

\section*{Model}
\begin{figure}[tb!]
\centering 
\includegraphics[width=\textwidth]{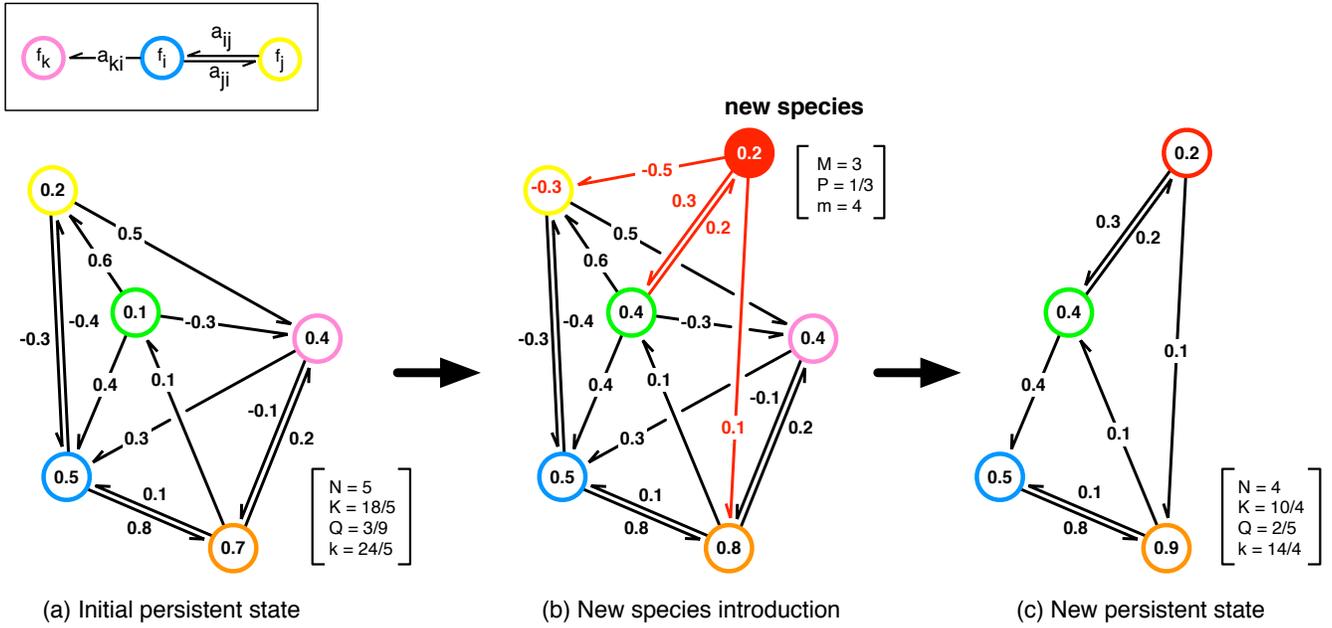}
\caption{
The temporal evolution of a partially bidirectional model.
(a): All the species have positive fitness and hence the system is in a persistent state.  
(b): The introduction of a new species (red). In this case the inclusion leads to the extinctions of a resident species (yellow) and that extinction triggers another extinction of a species, which is not directly connected to the newly introduced species (magenta). 
(c): Finally, the system reaches to a new persistent state.
The numbers in the square brackets demonstrate the control parameters; $M$: the number of interacting species of the newly introduced species, $P$: the ratio of the bidirectional relation in the initially assigned interactions, and $m=(1+P)M$: unidirectional link degree of the newly introduced species, and observables of the emerging network; $N$: the number of species, $K$: average number of interacting pairs per species in the system, $Q$: the ratio of the bidirectionally interacting pairs to the all interacting pairs in the system, and $k=(1+Q)K$: average unidirectional link degree of the system.}
\label{fig_model}
\end{figure}

As in this study we are interested in the effects of bidirectional influences on the robustness of the evolving open system, we take the previous model as the base describing it first in detail and then extend it by introducing bi-directionality in a controlled way. In the previously model the evolution of an open system of species was set up to take place in a graph or network of randomly connected species or links describing the influences between species. This dynamical system   evolves through interaction dependent survival of each species and a successive addition of new species to the system, as depicted in Fig.\ref{fig_model}. This network consists of nodes (species) that are sparsely connected to each other with directed and (positively or negatively) weighted links, each describing the strength of interaction between a pair of species. In this system the survival of each species is determined by its fitness, which is calculated by the sum of the weights of its incoming links such that if the fitness of a given species is positive, it will survive. On the other hand if the fitness of a species is non-positive that species and all the interactions from and to it are deleted. If two or more species have non-positive fitness values, the species with the minimum fitness are deleted and the fitnesses of other species are recalculate. This so called extinction process is repeated until all the remaining species have positive fitness and once such a network structure is reached, nothing further will happen until a new node is born or introduced into the system. According to the terminology of ecology this is called a {\it persistent} state rather than a stable state. 
\begin{table}
\begin{tabular}{|c|l|} \hline
	Control Parameters & $M$: number of interacting species for the new species\\
	(for newly introduced species) & $P$: ratio of the bidirectional relations\\
	& $m$: unidirectional link degree of the new species $\left(=(1+P)M \right)$\\
	\hline
	Observables & $K$: average number of interacting pairs per species\\
	(for emerging nets) & Q: ratio of the bidirectionally interacting pairs to the all interacting pairs  \\
	& $k$: average unidirectional link degree $\left( =(1+Q)K \right)$\\
	\hline
\end{tabular}
\caption{Parameters and observables of the present model}
\label{table_Parameters}
\end{table}
When the system has relaxed to such a persistent state, 
a new species (node) is added to the system, and the above described process continues. Next $M$ unidirectional links 
between the new node and the nodes randomly selected from the resident nodes are introduced. 

Till this point our extended model is the same as the previous model, but instead of introducing unidirectional links a mixture in total of $M$ unidirectional and bidirectional links is introduced by generating the bidirectional links with probability $P$ and unidirectional links with probability $1-P$. The weight for each link is chosen again from a normal distribution. Thus our extended model has only two control parameters, $M$ and $P$, instead of one in the previous model. In order to run the dynamics of both these models the above described addition and survival check process is repeated. A pseudo-code like description of the entire procedure of the present model is shown in the Methods section. 

Counting each bidirectional link as two unidirectional links,
one can say that $m = (1+P)M$ unidirectional links to the new species is introduced, which on average constitute $PM$ bidirectional links. The extreme cases of $P=0$ and $P=1$ correspond to the original unidirectional model (with the number of unidirectional links per new species, $m=M$) and the pure bidirectional model (with $m=2M$), respectively. We denote the density and the amount of bidirectionality of interactions in the emergent system in the same way, by the average number of interactions per species, i.e.  $K = \sum_i M_i/N$, where $M_i$ and $N$ represent the number of links of $\it{i}_{th}$ resident and total number of resident species, respectively. We denote the ratio of bidirectional links by $Q$. Then the resident species has $k=(1+Q)K$ unidirectional links such that they form $QK$ bidirectional links.
The above definitions are summarized in Tables 2.

\section*{Results}
\subsection*{Purely bidirectional interaction ($P = 1$)}
\subsubsection*{Transition in growth behavior at two-fold critical point}
\begin{figure}[tb!]
\centering 
\includegraphics[width=0.5\textwidth]{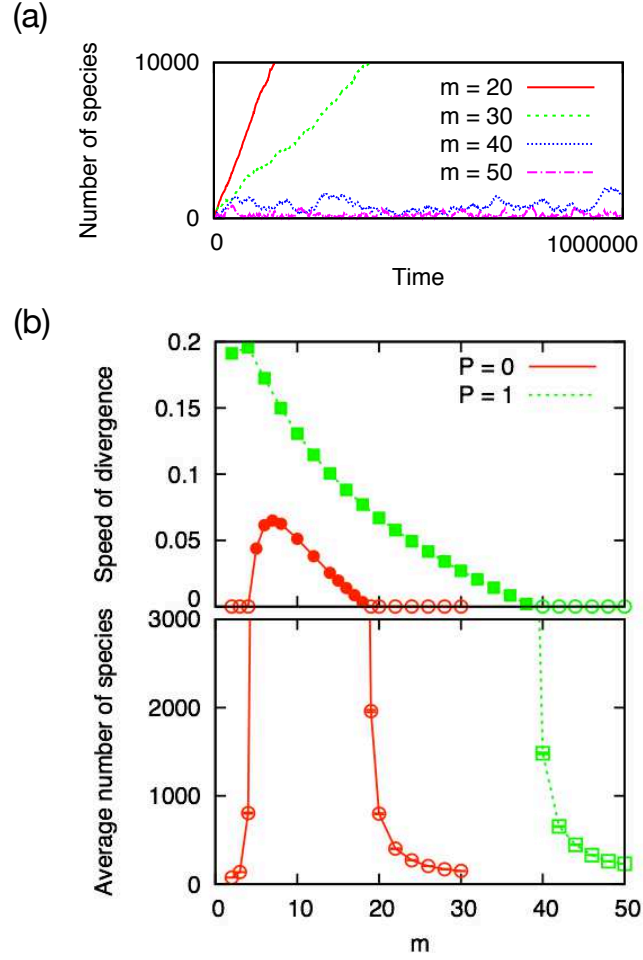}
\caption{
The critical behavior of emergent systems with $P=0$ or $P=1$. 
(a): Temporal evolution of total number of species $N(t)$ for $P=1$.
It diverges in time if the sum of in-degree and out-degree of each new species is small ($m = 2M = 20$ and $30$),
and it keeps fluctuating in a finite size for larger $M$ ($2M = 40$ and $50$).
(b): Phase diagram of the growth behavior for $P=0$ (red) and $p=1$ (green). 
(top) Speed of divergence. 
(bottom) Average number of species. 
Filled symbols represent the systems with positive speed $N(t)/t$. The bidirectional model (circles) involves a transition from the diverging phase to the finite phase between $m = 38$ and $40$. 
The critical point $m_C^{\rm{bi}} = 2M_C^{\rm{bi}} = 39$ is almost two-fold of that of the unidirectional model (triangles) $m_C^{P=0} = 18.5$. 
}
\label{fig_transition_behavior_bi}
\end{figure}

We first examine the system with purely bidirectional interaction, i.e. we choose $P=1$. Similarly with what was obtained for the model with unidirectional interactions, the purely bidirectional model yields two distinct phases characterized by its growth behavior. This is dependent on the unique parameter that specifies the number of interactions, $M$, we introduce to each new species(Fig.\ref{fig_transition_behavior_bi} (a)). While the system keeps growing without any limitation to its size if $M$ is smaller than a certain value $M_C$ ($\displaystyle \lim_{t \to \infty} N(t)/t > 0$), but the system shows finite size fluctuations if $M > M_C$ ($\lim_{t \to \infty} N(t)/t = 0$).
We call these phases diverging phase and finite phase, respectively.

From the systematic simulations, we have estimated the critical number of bidirectional interactions 
to be at between $M=19$ and $20$. As in this case all the interactions are bidirectional, the critical point, measured by the number of unidirectional links, is  $m^{P=1}_C = 2M^{P=1}_C = 39$ (Fig.\ref{fig_transition_behavior_bi} (b)). This value is almost two times that of the pure unidirectional model (i.e. $m^{P=0}_C = 18.5$). 
Another clear difference from the unidirectional model is the absence of a finite phase in the very sparse regime. In the unidirectional case, the systems with $m \le 4$ are in finite phase, due to the fragility of the emerging network. In contrast, in the bidirectional case, the system even with $M=1$ turns out to exhibit diverging behavior. 
This is basically caused by the easiness of forming mutually supporting ``dimers'', which is enough to let the nodes survive even when they are not globally connected. This regime will be discussed later in more detail. Our primary aim in what follows is to understand the two-fold transition point in the well-connected regime.

To illustrate why we measure the interaction density by the number of unidirectional links and call the transition point ``two fold'', we will 
briefly review the mean field argument on the basis of which the understanding and the rough prediction of the transition in the unidirectional model was obtained\cite{Shimada2014SREP,Shimada2016}.
In a system the introduction of a new species can lead to the extinctions of the resident species. 
We denote the extinction probabilities during this event as $E_i$. It should be noted that the extinction of resident species can also cause a cascading extinction through the deletion of the links from it, with a certain probability $E_e$. Assuming a structure-less random network topology, which is supported by the direct observation of the emergent network, the average number of extinctions $N_E$ per a newly introduced species can be calculated as a simple infinite geometric series as follows
\begin{equation}
N_E^{P=0} = \frac{mE_i}{2-kE_e}
\label{eq_ne_uni}
\end{equation}
for the unidirectional model, and, 
\begin{equation} 
N_E^{P=1} = 2ME_i \left(\frac{1+E_e}{2-2KE_e+2E_e} \right)
\label{eq_ne_bi}
\end{equation}
for the bidirectional model. Here the critical point corresponds to $N_E = 1$. Substituting $E_i$ and $E_e$ by an average extinction probability $E$ and $k$ by $m$, we obtain a simple guide of the transition point, namely $mE = 1$. Therefore we expect $m_C^{P=1} E = 2 M_C^{P=1} E =1$ for the bidirectional model, which means that the critical point of the bidirectional model should have the same number of unidirectional links $m_C^{P=1} = 2M_C^{P=1}$ with that of the unidirectional model $m_C^{P=0} = 18.5$ \cite{Shimada2014SREP}. The details and an enhanced version of this mean-field argument are shown in the Methods section.

\subsubsection*{Reinforcement of elements}
\begin{figure}[tb!]
\centering 
\includegraphics[width=0.5\textwidth]{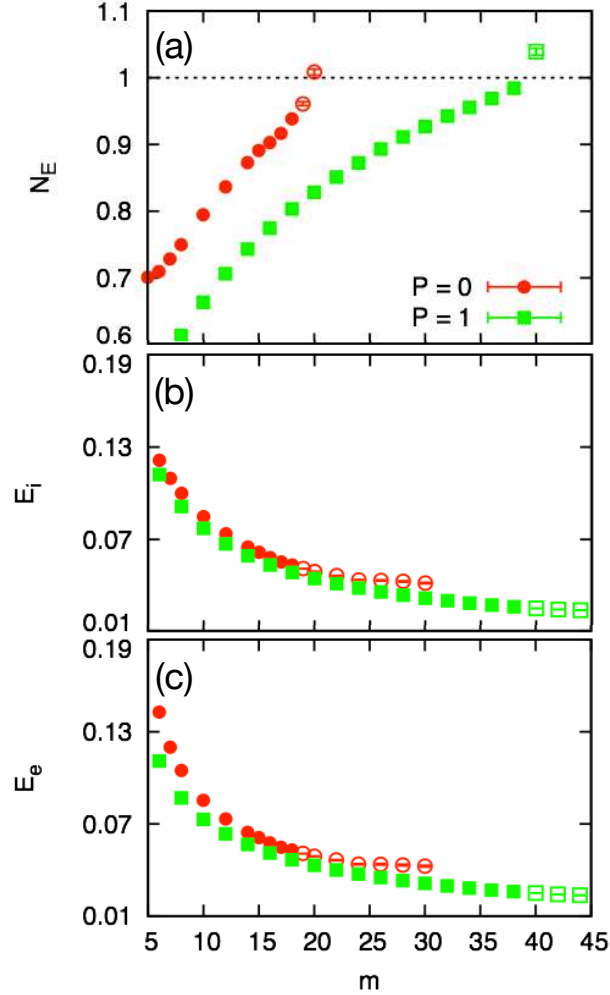}
\caption{
The mean-field approximation predictions of the transition points.
(a): Estimation of the transition point from the diverging phase to the finite phase. The average number of total extinctions, $N_E^{P=0}$ (circles) and $N_E^{P=1}$ (boxes), increases monotonically as a function of $m$. They cross the critical value 1 at $m^* = 19.5$ for the unidirectional model and at $2M^* = 39$ for the bidirectional model. Both of $m^*$ and $2M^*$ agree well with the real transition points $m_C^{P=0} = 18.5$ and $2M_C^{P=1} = 39$. 
(b): Extinction probability $E_i$. (c): Extinction probability $E_e$.
For the unidirectional model (circles) and the bidirectional model (boxes) with the same unidirectional link degree of a newly incoming species $m$, the extinction probabilities $E_i$ and $E_e$ of the bidirectional model are smaller than the corresponding probabilities of the unidirectional model.
}
\label{fig_meanfield_test}
\end{figure}

In order to clarify the crucial factor for the shift of the transition point, 
we evaluate the transition point using Eq.(\ref{eq_ne_uni}) and Eq.(\ref{eq_ne_bi}) with empirically obtained extinction probabilities (see the Method section). 
The average number of total extinctions per a newly introduced species increases monotonically as a function of the degree of a new species both for the unidirectional model and for the bidirectional model. The quantities $N_E^{P=0}$ and $N_E^{P=1}$ cross the critical value 1 at $m^* \sim 19.5$ for the unidirectional model and $2M^* \sim 39$ for the bidirectional model, respectively, as depicted in Fig.\ref{fig_meanfield_test} (a). These critical points $m^*$ and $2M^*$ agree well with the real transition points, $m_C^{P=0} = 18.5$ and $2M_C^{P=1} = 39$.

This result suggests that the shift of the transition point stems from the decrease in the extinction probability of the emergent system. In Fig.\ref{fig_meanfield_test} (b) we see that the quantities $E_i$ and $E_e$ decrease monotonically as a function of the degree of a new species both for the unidirectional and bidirectional models. Comparing the bidirectional model with the unidirectional model, however, the bidirectional model gives a smaller $E_i$ and $E_e$ if the new species of the unidirectional model and the bidirectional model are assigned same number of unidirectional links, $m^{P=0} = 2M^{P=1}$. 
Further investigation of the emerging system reveals that the larger decrease in the extinction probability of the bidirectional model is caused by the slight increase in the average degree, together with the small change in the link weight correlation (see Method section for more details).
 
\subsection*{Model with partially bidirectional interaction ($P \in [0,1]$)}
\subsubsection*{Bidirectionality dependence of the transition of the growth behavior}
\begin{figure}[tb!]
\centering 
\includegraphics[width=\textwidth]{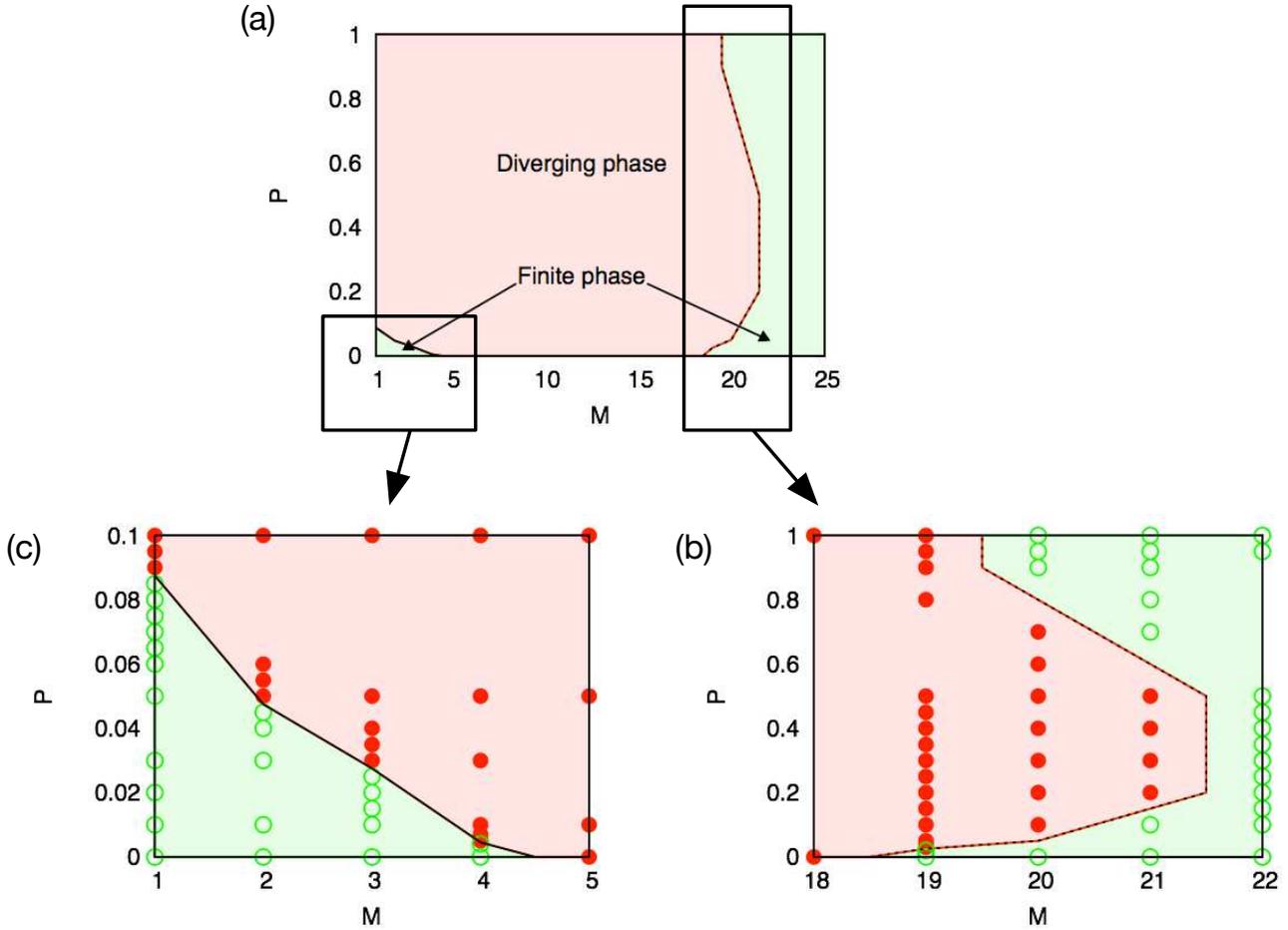}
\caption{
The phase diagram of the growth behavior for (a) $1 \leq M \leq 22$, (b) the dense interaction regime $18 \leq M \leq 22$, and (c) the sparse interaction regime $1 \leq M \leq 5$. The partial bidirectionality expands the area of the diverging phase (filled circles). There exists an upper phase boundary (dotted line) at around $18 < M < 22$ and a lower phase boundary (solid line) at around $1 \leq M < 5$. The upper boundary is asymmetric for the bidirectionality and there is an optimum bidirectionality for the growth rate at around $P \sim 0.3$. 
The finite phase (open circles) in the sparse connecting regime $M < 5$ remains within a range of small but finite bidirectionality. The area of the finite phase gradually decreases as a function of the increasing bidirectionality.
}
\label{fig_phase_diagram_partially}
\end{figure}

Next, we examine the systems with partially bidirectional interactions to investigate the influence of the proportion of bidirectionality $P$ on the growth behavior. 
We find that $P \in (0,1)$ expands the area of the diverging phase, as seen in Fig.\ref{fig_phase_diagram_partially} (a). Interestingly or even surprisingly, the system with a certain level of bidirectionality steadily grows even if the system with purely unidirectional interactions ($P = 0$) or purely bidirectional interactions ($P = 1$) is in the finite phase. There are two phase boundaries, one found at around $18 < M < 22$ and another at around $1 \leq M < 5 $, as can be seen in Fig.\ref{fig_phase_diagram_partially} (a). Here the finite phase remains not only in the dense interaction regime and but also in the sparse interaction regime. 

One may expect a strong effect of birectionality, because strong accumulation and cooperative interaction is possible.  The most prominent example of this is the case of $M=1$, in which each cooperative dimer can survive independently letting the system to consist of many independent clusters with broad size distribution. However, in the dense regime, one cannot find such a strong structuring effect. For example, in the dense regime, the proportion of bidirectional links in the emerging system $Q$ remains almost the same as the initial proportion $P$ (see Methods).
 
\subsubsection*{Growth behavior in the dense interaction regime ($ 19 \leq M \leq 22$)}
Figure \ref{fig_growth_behavior_partially} (a) shows the speed of divergence in the dense interaction regime. Here the speed of divergence is asymmetrically distributed as a function of a proportion of the bidirectional links. It is seen that there is an optimum bidirectionality for the growth rate $P \sim 1/3$. Thus, the upper phase boundary in the denser interaction regime ($18 < M < 22$) is asymmetric for the proportion of the bidirectional links (Fig.\ref{fig_phase_diagram_partially} (b)). We also find that there exists an optimum bidirectionality for the growth rate at around $P \sim 1/3$. 

The system with a moderate amount of bidirectionality grows for $M = 19, 20$, and $21$, even if the system with purely unidirectional interactions or purely bidirectional interactions ($P=0$ or $P=1$) is in the finite phase. In addition, a small amount of bidirectionality makes the partially bidirectional system grow. On the other hand, the system with $M = 22$ remains in the finite phase only for all $P \in [0,1]$. 

\subsubsection*{Growth behavior in the sparse interaction regime ($1 \leq M \leq 4$)}
\begin{figure}[tb!]
\centering
\includegraphics[width=\textwidth]{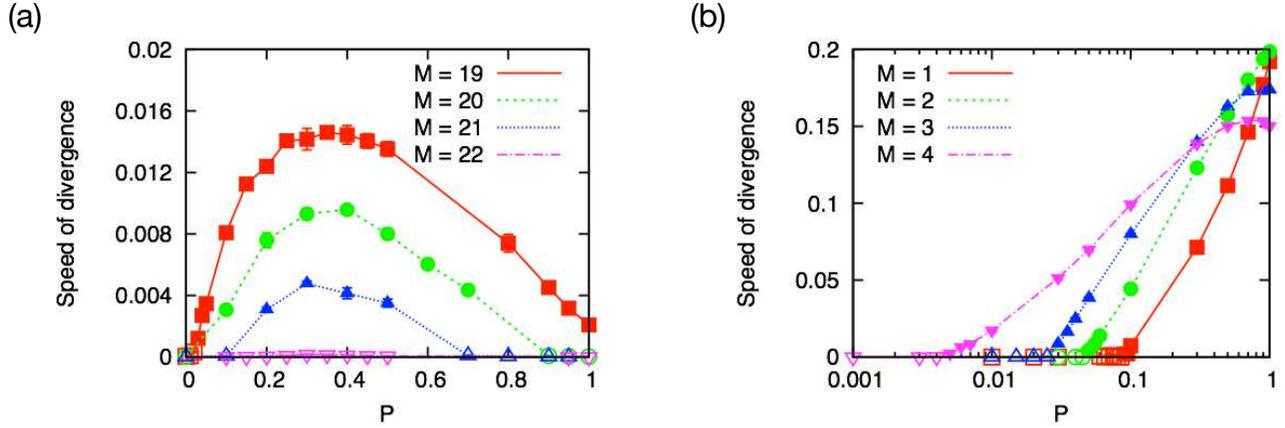}
\caption{
The growth behavior of the partially bidirectional systems in
(a) the dense interaction regime and (b) the sparse interaction regime. 
(a): Moderate proportion of bidirectional links brings the system from the finite phase to the diverging phase. The speed of divergence is asymmetric as a function of the bidirectionality and there exists an optimum bidirectionality for the growth rate at around $P \sim 1/3$. 
(b): The speed of divergence increases monotonically from zero as a function of proportion of the bidirectional links. A small but finite bidirectionality brings the sparsely connected system from the finite phase to the diverging phase. 
}
\label{fig_growth_behavior_partially}
\end{figure}

In Fig.\ref{fig_growth_behavior_partially} (b) it is seen that the speed of divergence increases monotonically from zero with the increasing proportion of the bidirectional links. Thus, a small but finite amount of bidirectionality brings the system from the finite phase to the diverging phase, as is evident in Fig.\ref{fig_phase_diagram_partially} (c). On the other hand, in this regime the area of the finite phase gradually decreases as a function of the amount of bidirectionality. Then if the system has larger bidirectionality than a certain value, the sparsely connected system involves only a diverging phase. 

\section*{Discussion}
We have studied the effects of introducing new species of variable degree of bidirectionality in the interactions between species focusing especially on the dynamics of the robustness of the network of species. 

It is first found that the system with purely bidirectional interactions is more robust against the inclusion of new species, which is characterized by the two-fold transition point from the growing phase to the finite phase.
The two fold transition point also means that one can roughly identify or even mix the unidirectional and bidirectional interactions to apply our framework to real systems, in which the detailed information of interactions and their nature is often not available.

In the model with partial bidirectionality, we find that there exists an optimum ratio of bidirectional interactions around $1/3$ for the growth of the system to take place. We also find that only a small, though finite, portion of bidirectional interactions let the very sparse ($M \le 4$) system to grow to a large and connected structure.

Except for the very sparse regime, all those drastic changes in the entire system's robustness are caused by a little change in the degree and also in the emergent correlations between the link weights. 
The network characteristics such as clustering and assortativity are also found to be not much deviated from that of Erd\H{o}s-R\'{e}nyi random graph (see Methods).
This is quite noteworthy because the system in our model self-emerges without any apparent constraints and hence more drastic self-organization could be possible.
A study of an extended model in which we force the system to have stronger correlations in the link and weight topologies could be the next step, to further relate our framework to the real systems\cite{Kondoh2003, JANE:JANE1460, Pimm1991, MougiKondoh2012, CoyteSchluterFoster2015, Tokita2015SREP}.

\newpage

\section*{Methods}
\subsection*{A pseudo-code representation of the model}
The present model is expressed as the following pseudo-code:
\begin{enumerate}
	\setcounter{enumi}{-1}
	\def\theenumii{\roman{enumii}}
	\item (Create an initial network.)
    
	\item Calculate the fitness for each species: $\displaystyle f_{i} = \sum_j^{incoming} a_{ij}$
	
	\item If the fitness of the species are all positive (Fig. \ref{fig_model} (a) and (c), persistent state), go to step 3. If not,
		\begin{enumerate}
			\item Delete the species with minimum (and non-positive) fitness value 
			\item Delete the links connecting to and from that species.
			\item Go back to the step 1 (to re-evaluate the fitnesses).
		\end{enumerate}
	
	\item A new species is added to the system (Fig. \ref{fig_model} (b)).
		\begin{enumerate}  
			\item $M$ interacting species are chosen randomly from the resident species, with equal probability $1/N(t)$.
			\item The new species forms bidirectional links, with probability $P$. 
            \item With probability $1-P$, we assign a unidirectional link. Its direction is randomly chosen with probability $1/2$.	
			\item The link weights are independently drawn from the standard normal distribution. The link weights between incoming and outgoing unidirectional links which consist up one bidirectional link are also independent: $\langle a_{ij}a_{ji} \rangle = 0$.
            \item Go to 1.
		\end{enumerate}
\end{enumerate}

\subsection*{The mean-field estimation of the transition point}
In the original unidirectional model ($P=0$), the system keeps growing only if the interaction is moderately sparse ($5 \leq m \leq 18$) and cannot grow outside of this regime ($m \leq 4$ and $m \geq 19$).
The basic mechanism of the transition from the diverging phase to the finite phase at $m_c = 18.5$ is known to be explained by the mean-field argument \cite{Shimada2014SREP, Shimada2015MABS}, which assumes a correlation-less random network structure for the interaction network of self-emerging system. The validity of this assumption is confirmed  in the simulation, from the direct observation of the system.
With such an assumption, the system's growth behaviour against the inclusions of new species is characterized by two average extinction probabilities, $E_i$ and $E_e$. $E_i$ is the extinction probability of the resident species after obtaining a direct incoming link from the newly introduced species. The other extinction probability, $E_e$, is the probability of a species to get extinct after loosing one incoming link which is brought by the extinction of the neighboring species.
Because of the assumption of random network topology, the total number of extinctions of the resident species per newly introduced species $\langle N_e \rangle$ is calculated from a simple geometric sum as
\begin{equation}
\langle N_e^{P=0} \rangle 
= \left( \frac{m}{2}E_i \right) \left[ \sum_{l=0}^{\infty}\left(\frac{k}{2}E_e\right)^{l} \right]
= mE_i\left(\frac{1}{2-kE_e}\right),
\label{eq_ne_uni_detail}
\end{equation}
where $mE_i/2$ is the average number of extinctions directly caused by the newly introduced species and the infinite geometric sum corresponds to the process of cascade extinctions. 
The transition point is the point at which $\langle N_E\rangle = 1$, i.e. the average number of gain and loss of the species are equal.

Let us perform the same calculation in the system which consists of fully bidirectional interactions ($P=1$).
In this case, the expected number of extinctions in the layer of direct connection to the newly introduced species is $ME_i$. These extinctions cause the following link-deletion events, subsequently giving rise to $KE_e$ extinctions in the next-nearest neighbors. After this layer, the contribution of the cascading extinction in each deeper layer is $(K-1)E_e$.
Therefore, the total number of extinctions per a newly introduced species in the bidirectional model is calculated as,  
\begin{equation}
\langle N_e^{P=1} \rangle 
= ME_i \left[ 1 + KE_e\sum_{l=0}^{\infty}\left((K-1)E_e \right)^{l} \right] 
= 2ME_i \left(\frac{1+E_e}{2-2KE_e+2E_e} \right).
\label{eq_ne_bi_detail}
\end{equation}

For the simple evaluation of the transition point, we further assume that the average degree of the emergent system is same as the initial degree, i.e. $m = k$ and $M = K$.
Because the correlation-less network assumption it also means that the extinction probability is a function of the unique system parameter, i.e. average degree, meaning that it can be written as a function of initial degree $m=2M$. Based on the empirical findings, we also assume that the extinction probabilities are similar and much smaller than $1$ near the transition point ($E_c = E_i = E_e \ll 1$).
Then the condition for the unidirectional case (Eq. \ref{eq_ne_uni_detail}) becomes the following very simple form \cite{Shimada2014SREP}
\begin{equation}
 m_c \cdot E(m_c) = 1,
\end{equation}
and the corresponding simple form for the bidirectional system (Eq. \ref{eq_ne_bi_detail}) 
is 
\begin{equation}
(2M_c) \cdot E(2M_c) = 1 + \frac{E_c}{2+E_c} \sim 1.
\end{equation}
These results gives us the basic estimation for the transition point for the bidirectional systems, which says that the transition occurs at the same critical unidirectional degree: 
\begin{equation}
	2M_c = m^{P=1}_c = m^{P=0}_c.
\end{equation}
This is why we regard the observed transition point, $M^{P=1}_c \sim m^{P=0}_c$,
``tow-fold'' of what is naively expected.

\subsection*{Calculation of the extinction probabilities in the emerging systems}
The extinction probabilities $E_i$ and $E_e$ in an emerging system are calculated according to those definitions as,
\begin{equation}
E_i = \frac{1}{N}\sum_{i=1}^{N}\left(\frac{1}{\sqrt{2\pi}}\int_{-\infty}^{-f_i}\exp\left(-\frac{a^2}{2}\right)da\right), 
\label{eq_ei}
\end{equation}
\begin{equation}
E_e = \frac{\displaystyle \#_{a_{ij} \leq f_i}}{\displaystyle \#_{link}}, 
\label{eq_ee}
\end{equation}
where $\#_{a_{ij} \leq f_i}$ and $\#_{link}$ represent the number of species with smaller fitness than the link weight receiving from each link and the total number of links of the emergent system, respectively. 

\begin{figure}[tb!]
\centering 
\includegraphics[width=\textwidth]{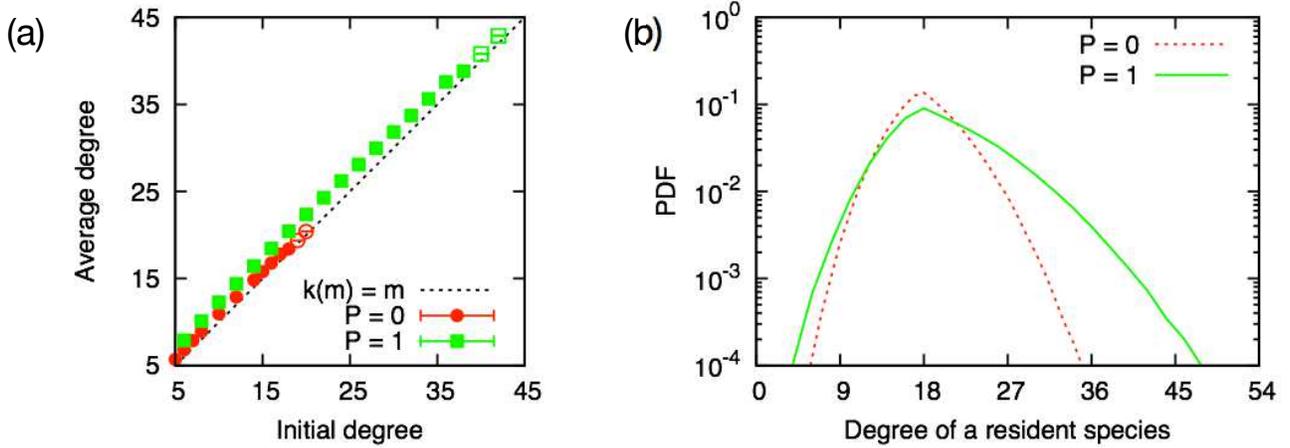}
\caption{
(a): Average degree of the emergent system. The emergent system in the bidirectional model (boxes) has larger average degree than that of the unidirectional model (circles) with the same unidirectional link degree of the new species $m^{P=0} = 2M^{P=1}$.
(b): Degree distribution of the emergent system at $m = 18$. The degree distribution of the bidirectional model (solid line) has broader tails than that of the unidirectional model (dashed line), with sharing the same peak position. 
}
\label{fig_degree}
\end{figure} 
\begin{figure}[tb!]
\centering 
\rotatebox{0}{\includegraphics[width=\textwidth]{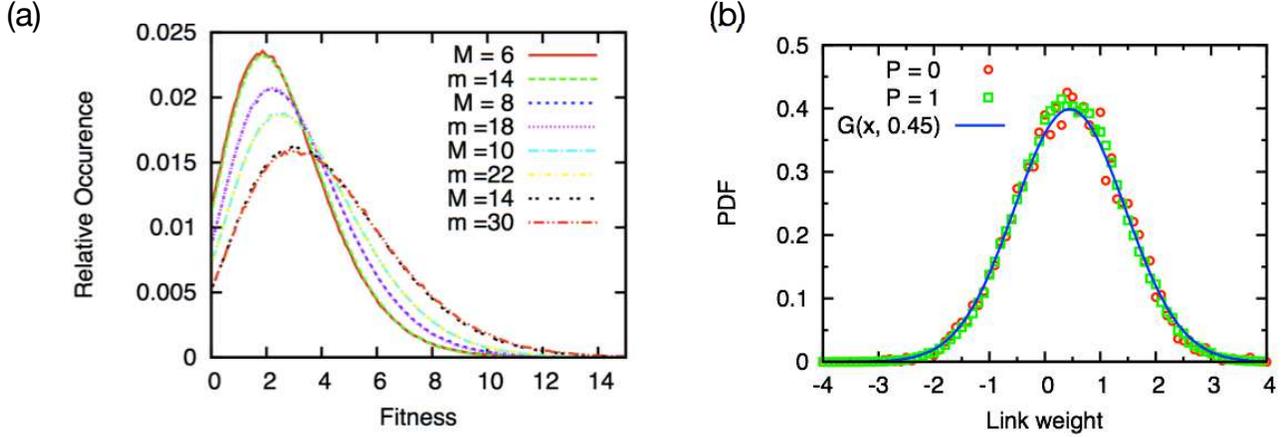}}
\caption{
Scaling relation in the fitness distribution function. 
(a): Fitness distributions of the unidirectional model ($m= 14, 18, 22$, and $30$) and of the bidirectional model ($M=6, 8, 10$, and $14$).
(b): Link weight distributions of the unidirectional model with $m^{P=0} = 18$ and  of the bidirectional model with $M^{P=1} = 8$.
Solid blue line is a fitting function, a standard normal distribution function $G(x,\mu)$ (solid line) with its mean $\mu = 0.45$.
}
\label{fig_fitness_scaling}
\end{figure}

\subsection*{The network structure and its effect in the bidirectional model ($P=1$)}
\subsubsection*{Degree distribution}
As shown in the Fig. \ref{fig_degree} (a), the average unidirectional link degree of the emergent system $k$ increases almost linearly as a function of the degree of a new species $m$, both for the unidirectional model and for the bidirectional model.
For the unidirectional model and the bidirectional model with the same degree of a newly incoming species $m^{P=0} = 2M^{P=1}$, the average degree of the bidirectional model is slightly larger than that of the unidirectional model. 
This difference comes mainly from the broader tail of degree distribution in the bidirectional case. The peak position of the degree distribution at $k = m$ and the tail on the large degree side is fatter than that on the other side (Fig.\ref{fig_degree} (b)). 

One can find a scaling relation in the fitness distributions of the unidirectional model and the bidirectional model (Fig.\ref{fig_fitness_scaling} (a)). 
The fitness distributions of
the unidirectional model and the bidirectional model match well when the control parameter of those satisfies the relation: $m^{P=0} = 2(M^{P=1}+1)$.
This scaling relation holds also for the distributions of the link weights (Fig.\ref{fig_fitness_scaling} (b)). All the results above can be explained by the slight increase in the actual average degree in the emergent system in the bidirectional case:
$k^{P=1}(M) = k^{P=0}(2(M+1))$.

\begin{figure}[tb!]
\centering 
\includegraphics[width=\textwidth]{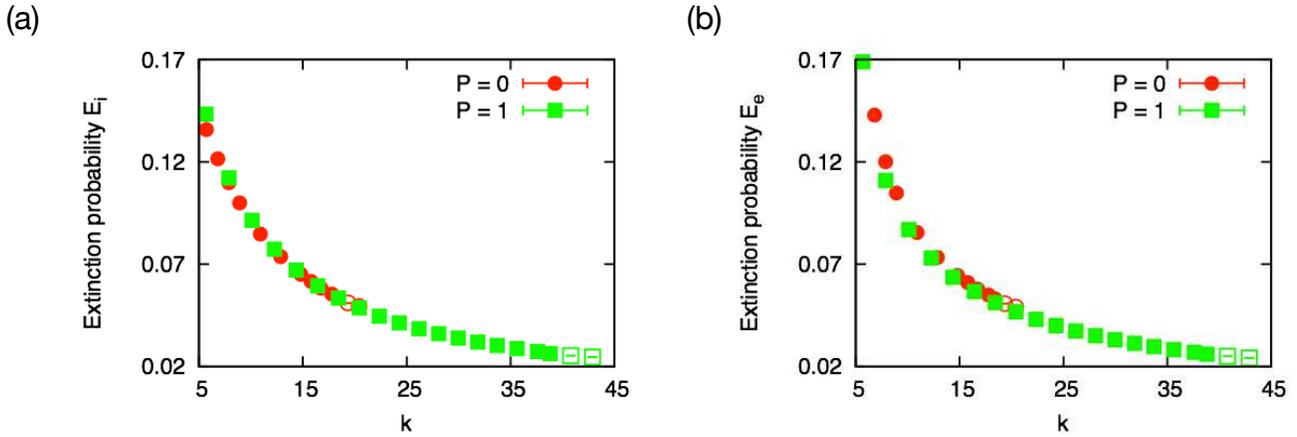}
\caption{
(a): Extinction probability $E_i$. 
(b): Extinction probability $E_e$. 
Emergent systems of the unidirectional and bidirectional models with same average unidirectional link degree show the same extinction probability against the direct disturbance event from the new species, $E_i$. On the other hand, the extinction probability against the loss of a link, $E_e$, is slightly smaller for the bidirectional model than that of the unidirectional model.
}
\label{fig_extinction_probability_scaling}
\end{figure}

\subsubsection*{Weak correlation in link weights}
According to the scaling relation, the unidirectional model and the bidirectional model give the same average degree of the emergent systems 
$k^{P=0} = 2K^{P=1}$ at $m^{P=0} = 2(M^{P=1}+1)$. 
Therefore, the unidirectional model and the bidirectional model are expected to have the same extinction probabilities at $k^{P=0} = 2K^{P=1}$ (Fig.\ref{fig_extinction_probability_scaling} (a)). 
However, while $E_i$ is found to obey this relation, $E_e$ of the bidirectional model is slightly smaller than that of the unidirectional model (Fig.\ref{fig_extinction_probability_scaling} (b)).
It suggests that the system of the bidirectional model has stronger weight-fitness correlation than the unidirectional model.

Here we conclude that the shift of the transition point stems from the small decrease in the extinction probability of the bidirectional model due to the weak shift in the average degree and the correlation in link weights, not from strong structuring or emergence of a certain motif in the emergent system. 

\subsection*{The network structure of the partially bidirectional model}

\subsubsection*{Dense regime}
The self-emerging network of the partially bidirectional model in the dense regime is found to have only weak structuring, similar to the fully bidirectional case.
The proportion of bidirectional links in the system also remains the same as the newly introduced interactions ($Q=P$). 
\subsubsection*{Sparse regime}
In the sparse regime, the emergent systems tend to have more of bidirectional links than that of newly introduced links ($Q > P$). This indicates the relatively strong self organization effect in this regime, which is regarded as the main reason for the vanish of the finite phase only for small values of $P$. 

\begin{figure}[tb!]
\centering 
\includegraphics[width=\textwidth]{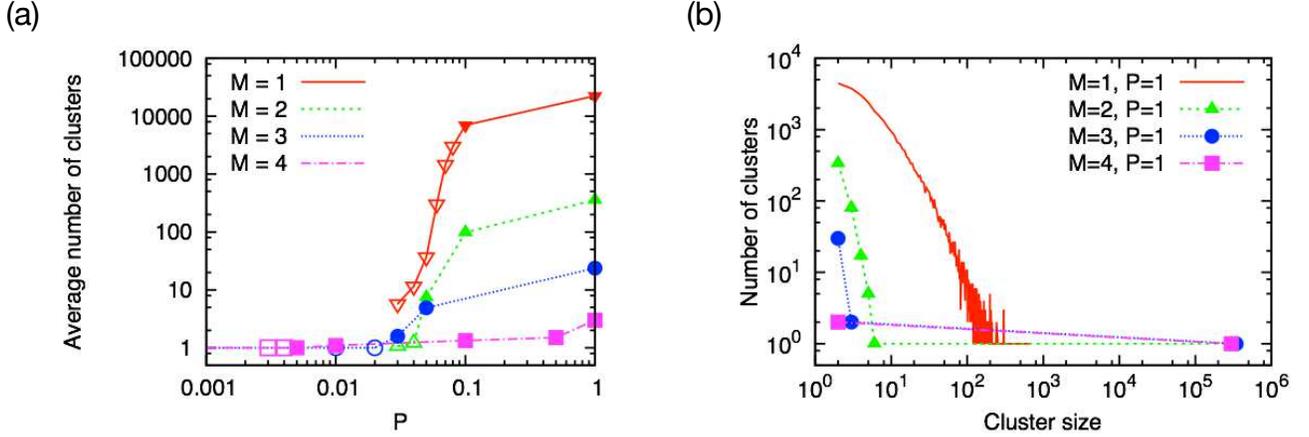}
\caption{
(a): Average number of clusters of the emergent systems at $t = 9.5 \times 10^6$. Except for the case of $M=1$, The systems in the finite phase (open symbols) have single cluster, i.e., those are connected. The systems in the diverging phase (filled symbols) consist of multiple independent clusters. 
(b): Cluster size distributions of the systems with purely bidirectional interactions in the diverging phase, at the same time step ($t = 9.5 \times 10^6$).
In the case of $M = 1$ (open reserve triangles), the distribution is continuous and broad up to the largest cluster size. 
Systems with $M > 1$ are composed of a single big cluster and some small fragments.
}
\label{fig_network_structure_sparse}
\end{figure}

In order to investigate the network structure in this regime, we analyze the network size distribution. 
As shown in Fig.\ref{fig_network_structure_sparse} (a), the systems are composed of multiple clusters in the diverging phase, whereas all resident species construct a single big cluster in the finite phase.
In the diverging phase, in contrast, the systems with $M = 2, 3$ and $4$ are composed of a single big cluster with some small fragments such as dimers, trimers, etc. (Fig.\ref{fig_network_structure_sparse} (b)). The cluster size distribution of the system with $M = 1$ is continuous up to its largest cluster size, meaning that there is no dominant giant component. 

The basic mechanism of the self-organization seems to be a formation of cooperative dimers. Such a dimer can survive independently and hence can be a nuclei of non-connecting cluster. We can see this effect in the cluster size distribution clearly in the case of $M=1$. Although it becomes less visible in the cluster size distribution in the case of $M>1$, because of the random reconnection (or, bridging) process due to the introduction of new nodes, the dimer nucleus still let the network, which consists of sparsely connected almost independent clusters, survive.
\color{black}

\subsection*{Assortativity, Nestedness, and Clustering}
Some network characteristics of the emergent systems are shown in Table \ref{table_network_characteristics}.
For this we treat unidirectional and bidirectional links equally as directionless links. Therefore the assortativity coefficient is calculated from the directionless link degrees of the nodes in the both ends of link $i$,
$\kappa^\alpha_i$ and $\kappa^\beta_i$, as follows
\begin{equation}
	A = \frac{\sum_i (\kappa_i^\alpha - \langle \kappa^\alpha \rangle)(\kappa_i^\beta - \langle \kappa^\beta \rangle)}
	{\sqrt{\sum_i (\kappa_i^\alpha - \langle \kappa^\alpha \rangle )^2}\sqrt{\sum_i (\kappa_i^\beta - \langle \kappa^\beta \rangle)^2}}.
\end{equation}
Moreover, we adopt a nestedness measure for uniparpite and dilectionless networks\cite{Lee2012} as
\begin{equation}
	B =
	\frac{2}{N(N-1)} \sum_{i}^{N-1} \sum_{j > i}^N
	\left( \frac{\displaystyle \sum_{l=1}^N a_{il} a_{jl}}{\displaystyle {\rm min} (K_i, K_j)} \right),
\end{equation}
where $a_{ij}$ and $K_i$ represent the adjacency matrix element and the degree of species $i$, defined by the directionless links.
The clustering coefficient is also calculated from the number of triples and triangles formed by the directionless links as,
\begin{equation}
	C = \frac{3 \times \mbox{(number of triangles)}}{\mbox{number of triples}}.
\end{equation}
Comparing to the Erd\H{o}s-R\'{e}nyi random graph, clustering coefficient is slightly smaller and nestedness is almost the same.
In addition the assortativity coefficient turns out to be slightly negative.
\begin{table}[h]
\centering 
\begin{tabular}{|c|c||c|c|c||c|l|l|}
	\hline
    $M$ & $P$ & $\langle N \rangle$ & $\langle K \rangle$ &$\langle Q \rangle$ & assortativity coefficient& nestedness & clustering coefficient\\
	\hline
       & 1   & $N_{obs} =$ 20,000 & 19.7 & 1.00  & -0.0202 & 0.00262 (1.09) & 0.00221 (0.917)\\
       & 0.7 & $N_{obs}$ & 19.7 & 0.703 & -0.0214 & 0.00107 (1.09) & 0.000904 (0.917)\\
       & 0.5 & $N_{obs}$ & 19.8 & 0.504 & -0.0216 & 0.00107 (1.09) & 0.000921 (0.931)\\
    19 & 0.3 & $N_{obs}$ & 19.9 & 0.304 & -0.0223 & 0.00108 (1.09) & 0.000907 (0.913)\\
       & 0.1 & $N_{obs}$ & 19.7 & 0.102 & -0.0210 & 0.00107 (1.09) & 0.000901 (0.916)\\
       & 0   & $2.67 \times 10^3$ & 19.4 & 0.00  & -0.0198 & 0.0103 (0.841) & 0.00866 (0.917)\\
	\hline
    10 & 0 & $N_{obs}$ & 10.9 & 0.00 & -0.0360 & 0.000610 (1.12) & 0.000469 (0.861)\\
	\hline
       & 1     & $N_{obs}$ & 5.04 & 1.00    & -0.0444 & 0.000305 (1.21) & 0.000205 (0.814)\\
     4 & 0.1   & $N_{obs}$ & 5.00 & 0.116   & -0.0590 & 0.000293 (1.18) & 0.000233 (0.942)\\
       & 0.004 & $7.17 \times 10^3$  & 4.58 & 0.00540 & -0.0763 & 0.00130 (0.911) & 0.000745 (0.658)\\
	\hline
       & 1    & $N_{obs}$ & 3.93 & 1.00   & -0.0512 & 0.000245 (1.24) & 0.000171 (0.869)\\
     3 & 0.1  & $N_{obs}$ & 3.87 & 0.126  & -0.0765 & 0.000231 (1.20) & 0.0000970 (0.501)\\
       & 0.02 & $1.35 \times 10^3$  & 3.70 & 0.0305 & -0.0836 & 0.00332 (0.878) & 0.00169 (0.603)\\
	\hline
	   & 1    & $N_{obs}$ & 2.84 & 1.00   & -0.0761 & 0.000182 (1.28) & 0.0000855 (0.601)\\
	 2 & 0.1  & $N_{obs}$ & 2.81 & 0.153  & -0.0973 & 0.000172 (1.22) & 0.0000479 (0.341)\\
	   & 0.04 & $1.16 \times 10^3$ & 2.80 & 0.0726 & -0.0900 & 0.00241 (1.00)  & 0.000817 (0.418)\\
	\hline
 	   & 1    & $N_{obs}$ & 1.84 & 1.00  & -0.147 & 0.000120 (1.31) & 0.00 (0.00)\\
	   & 0.7  & $N_{obs}$ & 1.84 & 0.816 & -0.149 & 0.000118 (1.28) & 0.00 (0.00)\\
	   & 0.5  & $N_{obs}$ & 1.84 & 0.659 & -0.140 & 0.000117 (1.28) & 0.00 (0.00)\\
	 1 & 0.3  & $N_{obs}$ & 1.85 & 0.469 & -0.134 & 0.000118 (1.27) & 0.00 (0.00)\\
	   & 0.1  & $N_{obs}$ & 1.90 & 0.212 & -0.114 & 0.000125 (1.32) & 0.00 (0.00)\\
	   & 0.05 & $1.75 \times 10^3$ & 1.93 & 0.120 & -0.0788 & 0.00159 (1.14) & 0.00 (0.00)\\
	\hline
\end{tabular}
\caption{
Network characteristics of the emergent systems.
For the case in the diverging phase, we observe the system when it first reaches a certain size $N_{obs} = 20,000$.
The numbers in the parenthesis for clustering coefficient and nestedness are those ratios to the ones of Erd\H{o}s-R\'{e}nyi random graph with the same size and degree, i.e. $N = \langle N \rangle, K = \langle K \rangle, Q = \langle Q \rangle.$
}
\label{table_network_characteristics}
\end{table}

\bibliography{bidEOS}

\section*{Acknowledgements}
F.O. was partly supported by "Materials Research by Information Integration" Initiative (MI2I) project of the Support Program for Starting Up Innovation Hub from the Japan Science and Technology Agency (JST).
K.K. acknowledges financial support by the Academy of Finland Research project (COSDYN) No. 276439 and EU HORIZON 2020 FET Open RIA project (IBSEN) No. 662725.
JK was partially supported by H2020 FETPROACT-GSS CIMPLEX Grant No. 641191.
T.S. was partly supported by JSPS KAKENHI Grant Number 15K05202 and CREST, JST.

\section*{Author contributions statement}
F.O. and T.S. conceived the model and conducted the simulation. All authors analysed the results and wrote the manuscript.

\section*{Additional information}
Competing financial interests: The authors declare no competing financial interests.

\end{document}